\documentclass[5p, 10pt]{elsarticle}

\usepackage{amsmath,amssymb,graphicx,enumerate,multirow,array,ragged2e,siunitx}
\usepackage[colorlinks=true,breaklinks=true]{hyperref}
\makeatletter
\def\ps@pprintTitle{%
 \let\@oddhead\@empty
 \let\@evenhead\@empty
 \def\@oddfoot{}%
 \let\@evenfoot\@oddfoot}
\makeatother

\usepackage[caption=false]{subfig}

\biboptions{authoryear,round,sort}
\usepackage[dvipsnames]{xcolor}
\newcommand{\revision}[1]{{\color{black}{#1}}}

\newcommand{\revisiontwo}[1]{{\color{black}{#1}}}
\definecolor{revisiontwo}{rgb}{0,0,0}

\usepackage{framed} 
\usepackage{multicol} 
\usepackage{nomencl} 
\makenomenclature
\renewcommand*\nompreamble{\begin{multicols}{2}}
\renewcommand*\nompostamble{\end{multicols}}

\begin{document}

\begin{frontmatter}

\title{Rear-surface integral method for calculating thermal diffusivity from\\ laser flash experiments}

\author[qut]{Elliot J. Carr}
\ead{elliot.carr@qut.edu.au}

\address[qut]{School of Mathematical Sciences, Queensland University of Technology (QUT), Brisbane, Australia.}

\journal{Chemical Engineering Science}


\begin{abstract}
The laser flash method for measuring thermal diffusivity of solids involves subjecting the front face of a small sample to a heat pulse of radiant energy and recording the resulting temperature rise on the opposite (rear) surface. For the adiabatic case, the widely-used standard approach estimates the thermal diffusivity from the rear-surface temperature rise history by calculating the half rise time: the time required for the temperature rise to reach one half of its maximum value. In this article, we develop a novel alternative approach by expressing the thermal diffusivity exactly in terms of the area enclosed by the rear-surface temperature rise curve and the steady-state temperature over time. Approximating this integral numerically leads to a simple formula for the thermal diffusivity involving the rear-surface temperature rise history. \revision{Using synthetic experimental data we} demonstrate that \revision{the} new formula produces estimates of the thermal diffusivity -- for a typical test case -- that are more accurate and less variable than the standard approach. The article concludes by briefly commenting on extension of the new method to account for heat losses from the sample.
\end{abstract}


\end{frontmatter}

\section{Introduction}

\begin{table*}[!t]
   \begin{framed}
     \begin{multicols}{2}
     \revision{\textbf{Nomenclature}}\\
     \\
     \revision{\textit{Symbols}}\\
     \revision{\begin{tabular}{>{\RaggedRight\arraybackslash}p{0.8cm}>{\RaggedRight\arraybackslash}p{7.2cm}}
     $c$ & specific heat capacity of the sample [$\text{J}\,\text{kg}^{-1}\,\text{K}^{-1}$]\\
     $h_{0}$ & scaled heat transfer coefficient at the front surface of the sample [$\text{m}^{-1}$]\\
     $h_{L}$ & scaled heat transfer coefficient at the rear surface of the sample [$\text{m}^{-1}$]\\
     $k$ & thermal conductivity of the sample [$\text{W}\,\text{m}^{-1}\,\text{K}^{-1}$]\\     
     $\ell$ & depth of the layer at the front surface of the sample in which the heat pulse is instantaneously absorbed [$\text{m}$]\\     
     $L$ & length of the sample [$\text{m}$]\\
     \textcolor{revisiontwo}{$N$} & \textcolor{revisiontwo}{number of temperature samples recorded (excluding initial temperature) [--]}\\
     $Q$ & amount of heat absorbed through the front surface of the sample per unit area [$\text{J}\,\text{m}^{-2}$]\\  
     $T$ & temperature rise above the initial temperature of the sample after application of the heat pulse [$^\circ\text{C}$]\\
     $T_{r}$ & rear-surface temperature rise above the initial temperature of the sample after application of the heat pulse [$^\circ\text{C}$]\\
     \end{tabular}}   
     \vfill\null\columnbreak
     \vspace*{\baselineskip}
     \vspace*{\baselineskip}
     \revision{\begin{tabular}{>{\RaggedRight\arraybackslash}p{0.8cm}>{\RaggedRight\arraybackslash}p{7cm}}
     $T_{\infty}$ & value of $T_{r}$ at steady-state [$^\circ\text{C}$]\\
     $\widetilde{T}$ & synthetic rear-surface temperature rise data [$^\circ\text{C}$]\\
     $t$ & time [$\text{s}$]\\
     $t_{0.5}$ & half rise time [$\text{s}$]\\
     \textcolor{revisiontwo}{$t_{i}$} & \textcolor{revisiontwo}{$i$th sampled time [$\text{s}$]}\\
     \textcolor{revisiontwo}{$t_{N}$} & \textcolor{revisiontwo}{final sampled time [$\text{s}$]}\\
     \textcolor{revisiontwo}{$u(x)$} & \textcolor{revisiontwo}{$\int_{0}^{\infty}\left[T_{\infty} - T(x,t)\right]\,\text{d}t$ [$^\circ\text{C}\,\text{s}$]}\\ 
     $x$ & spatial coordinate [$\text{m}$]\\  
     $z$ & normally distributed random number [$^\circ\text{C}$]\\
     $\alpha$ & thermal diffusivity of the sample [$\text{m}^{2}\,\text{s}^{-1}$]\\
     $\widetilde{\alpha}$ & estimate of the thermal diffusivity of the sample [$\text{m}^{2}\,\text{s}^{-1}$]\\
     $\varepsilon$ & signed relative error [--]\\
     $\mu$ & mean\\
     $\rho$ & density of the sample [$\text{kg}\,\text{m}^{-3}$]\\
     $\sigma$ & standard deviation\\
     \textcolor{revisiontwo}{$\omega$} & \textcolor{revisiontwo}{dimensionless time [--]}
     \end{tabular}}   
     \end{multicols}
     \vspace*{-0.45cm}
   \end{framed}
\end{table*}

Originally proposed by \citet{parker_1961}, the laser flash method is the most popular technique for measuring the thermal diffusivity of solids \citep{vozar_2003,czel_2013,blumm_2002}. During the experiment, the front surface of a small sample of the solid is subjected to a heat pulse of radiant energy and the resulting temperature rise on the opposite (rear) surface recorded. In their classical paper, Parker et al. derived a formula  that allows the thermal diffusivity of the sample to be estimated from the half-rise time: the time required for the rear-surface temperature rise to reach one half of its maximum value. 

Parker et al.'s formula is derived under the assumption of ideal conditions \citep{vozar_2003,parker_1961,gembarovic_1994}: the heat flow is one-dimensional; the sample is homogeneous, isotropic, thermally insulated and of uniform thickness and uniform initial temperature; the heat pulse is instantaneously and uniformly absorbed by a thin layer of the sample at the front surface; and the density and thermo-physical properties of the sample are uniform, constant and invariant with temperature. Since Parker et al.'s initial paper, many modifications to the original method have been proposed for treating additional physical effects and configurations such as heat loss from the surfaces of the sample \citep{cowan_1963,heckman_1973}, finite pulse time effects \citep{cape_1963,heckman_1973,tao_2015,azumi_1981,taylor_1964} and layered samples \citep{czel_2013,chen_2010,james_1980,zhao_2016}. However, despite these improvements, the half-rise time approach remains standard practice under ideal conditions \citep{ASTM_E1461}.   

In this article, we present a novel way to calculate the thermal diffusivity from the rear-surface temperature rise history. In particular, we show how the thermal diffusivity can be expressed exactly in terms of the area enclosed by the rear-surface temperature rise curve and the steady state temperature over time. Approximating the integral representation of this area numerically leads to a new formula for the thermal diffusivity. \revision{Using synthetic experimental data we} demonstrate that \revision{the} new formula produces estimates of the thermal diffusivity -- for a typical test case -- that are more accurate and less variable than the standard half-rise time approach.

The remaining sections of this paper are arranged in the following way. In the next section, we briefly revisit Parker et al.'s original approach before presenting \revision{the} new approach for calculating the thermal diffusivity in Section \ref{sec:new}. \revision{Results} in Section \ref{sec:results} comparing \text{the} new formula to Parker et al.'s formula highlight the effectiveness of the new approach. The paper concludes in Section \ref{sec:conclusions} with a summary of the work and a discussion on extension of the new approach to the case of heat losses from the front and rear surfaces of the sample \citep{cowan_1963}.

\section{Parker et al.'s formula for thermal diffusivity}
\label{sec:parker}
We now briefly revisit the original approach of \citet{parker_1961} for calculating the thermal diffusivity from the half-rise time. Consider a sample of uniform thickness $L$ and \revision{let the thin layer in which the heat pulse is instantaneously and uniformly absorbed have depth $\ell$}. Under ideal conditions, the temperature rise, $T(x,t)$, above the initial temperature of the sample and after application of the heat pulse is governed by the following equations \citep{parker_1961}:
\begin{gather}
\label{eq:model_pde}
\frac{\partial T}{\partial t}(x,t) = \alpha\frac{\partial^{2}T}{\partial x^{2}}(x,t),\quad 0 < x < L,\quad t > 0,\\
\label{eq:model_ic}
T(x,0) = \begin{cases} \displaystyle\frac{Q}{\rho c\ell} & 0 < x <\ell\\ 0 & \ell < x < L,\end{cases}\\
\label{eq:model_bcs}
\frac{\partial T}{\partial x}(0,t) = 0,\quad\frac{\partial T}{\partial x}(L,t) = 0.
\end{gather}
where $\alpha$ is the thermal diffusivity of the sample, $Q$ is the amount of heat absorbed through the front surface per unit area, $\rho$ is the density of the sample, $c$ is the specific heat capacity of the sample, $x$ denotes spatial location and $t$ is time. Solving Eqs (\ref{eq:model_pde})--(\ref{eq:model_bcs}) for $T(x,t)$ and evaluating the resulting expression at the rear-surface, $x = L$, yields the theoretical rear-surface temperature rise curve \citep{parker_1961}:
\begin{gather}
\label{eq:Tr}
T_{r}(t) = T_{\infty}\left[1 + 2\sum_{n=1}^{\infty}(-1)^{n}\frac{\sin\left(n\pi\ell/L\right)}{n\pi\ell/L}e^{-n^{2}\omega(t)}\right],
\end{gather}
where $\omega$ is a dimensionless time and $T_{\infty}$ is the limiting steady state value of the rear-surface temperature rise:
\begin{gather}
\label{eq:dimensionless}
\omega(t) = \frac{\pi^{2}\alpha t}{L^{2}},\quad T_{\infty} = \frac{Q}{\rho c L}.
\end{gather} 
In their classical paper, \citet{parker_1961} arrive at a simple formula for the thermal diffusivity by arguing that $\sin(n\pi l/L)\approx n\pi\ell/L$ since $\ell$ is small. Using this approximation in Eq (\ref{eq:Tr}) yields:
\begin{gather}
\label{eq:rear_surface_temperature_rise_approx}
T_{r}(t)\approx T_{\infty}\left[1 + 2\sum_{n=1}^{\infty}(-1)^{n}e^{-n^{2}\omega(t)}\right] =: \widetilde{T}_{r}(t).
\end{gather}
The attraction of Eq (\ref{eq:rear_surface_temperature_rise_approx}) is that $\widetilde{T}_{r} = 0.5T_{\infty}$ when $\omega = \omega_{0.5} = 1.370$ (rounded to four significant figures) independently of the parameters in the model. The following approximate formula for the thermal diffusivity then follows from the expression for $\omega$ given in Eq (\ref{eq:dimensionless}):
\begin{align}
\label{eq:alpha_parker}
\alpha \approx \frac{\omega_{0.5}L^{2}}{\pi^{2}t_{0.5}},
\end{align}
where $t_{0.5}$ is the half rise time, that is, the time required for the rear-surface temperature rise to reach one half of its maximum value, $0.5T_{\infty}$. Equality in Eq (\ref{eq:alpha_parker}) is obtained in the limit $\ell\rightarrow 0$. We remark here that the constant 1.370 is originally reported as 1.38 by \citet{parker_1961}, however, as noted by several authors \citep{josell_1995,heckman_1973}, the actual value is 1.370 rounded to four significant figures. To ensure high accuracy, in this article we calculate $\omega_{0.5}$ in MATLAB (2017b) by solving $\widetilde{T}_{r}(\omega) = 0.5T_{\infty}$ using the bisection method, truncating the summation in Eq (\ref{eq:rear_surface_temperature_rise_approx}) at $n = 200$ terms, setting the approximation of $\omega_{0.5}$ at each iteration to be the midpoint of the bracketing interval and terminating the iterations when the width of the bracketing interval is less than $2\times 10^{-15}$. The resulting value of $\omega_{0.5}$ has a maximum absolute error of $10^{-15}$.


\section{New formula for thermal diffusivity}
\label{sec:new}
We now derive an alternative formula to Eq (\ref{eq:alpha_parker}) for calculating the thermal diffusivity from the rear-surface temperature rise. The approach involves obtaining a closed-form expression for the integral $\int_{0}^{\infty} \revision{\left[T_{\infty} - T(x,t)\right]}\,\text{d}t$, which provides the area enclosed by the rear-surface temperature rise curve, $T_{r}(t)$, and the steady state temperature, $T_{\infty}$, between $t = 0$ and $t\rightarrow\infty$. \revision{This idea is inspired by recent work on time scales of diffusion processes \citep{carr_2017,carr_2018,simpson_2013,ellery_2012b}, where closed-form expressions for similar integrals are given.}

Defining
\begin{gather}
\label{eq:u_adiabatic}
u(x) := \int_{0}^{\infty}\revision{\left[T_{\infty} - T(x,t)\right]}\,\text{d}t,
\end{gather}
and recalling that $T(L,t) = T_{r}(t)$, allows us to write:
\begin{gather}
\label{eq:u_integral_adiabatic}
\int_{0}^{\infty} \revision{\left[T_{\infty} - T_{r}(t)\right]}\,\text{d}t = u(L).
\end{gather}
To find $u(L)$ a boundary value problem satisfied by $u(x)$ is constructed. This is achieved by differentiating Eq (\ref{eq:u_adiabatic}) twice with respect to $x$ and utilizing Eqs (\ref{eq:model_pde}) and (\ref{eq:model_ic}) in the resulting equation \citep{carr_2017,carr_2018,simpson_2013,ellery_2012b}, yielding 
\begin{align}
\label{eq:u_ode_adiabatic}
u''(x) &=  \begin{cases} \displaystyle\frac{T_{\infty}L}{\alpha}\left[\frac{1}{\ell}-\frac{1}{L}\right],  & 0 < x <\ell,\\[0.3cm] \displaystyle-\frac{T_{\infty}}{\alpha}, & \ell < x < L.\end{cases}
\end{align}
Eq (\ref{eq:u_ode_adiabatic}) is supplemented by the following boundary and auxiliary conditions:
\begin{gather}
\label{eq:u_bcs_adiabatic}
u'(0) = 0,\quad u'(L) = 0,\\
\label{eq:u_continuity_adiabatic}
\text{$u(x)$ and $u'(x)$ are continuous at $x = \ell$},\\
\label{eq:u_constraint_adiabatic}
\int_{0}^{L} u(x) = 0.
\end{gather}
The boundary conditions, Eq (\ref{eq:u_bcs_adiabatic}), are derived by making use of the boundary conditions satisfied by $T(x,t)$, Eq (\ref{eq:model_bcs}), and noting that $u'(x) = \int_{0}^{\infty} - \frac{\partial T}{\partial x}(x,t)\,\text{d}t$ \citep{carr_2017,carr_2018}. The continuity conditions, Eq (\ref{eq:u_continuity_adiabatic}), follow from the definition of $u(x)$, Eq (\ref{eq:u_adiabatic}), and continuity of $T(x,t)$ and $\frac{\partial T}{\partial x}(x,t)$. Eq (\ref{eq:u_constraint_adiabatic}) is arrived at by considering:
\begin{gather*}
\int_{0}^{L} u(x)\,\text{d}x = \int_{0}^{L}\int_{0}^{\infty} \revision{\left[T_{\infty}-T(x,t)\right]}\,\text{d}t\,\text{d}x,
\end{gather*}
reversing the order of integration and noting that:
\begin{gather}
\label{eq:conservation}
\int_{0}^{L} T(x,t)\,\text{d}x = \int_{0}^{L} T_{\infty}\,\text{d}x.
\end{gather}
\revision{Eq (\ref{eq:conservation}) follows from the conservation implied by the thermally-insulated boundary conditions \citep{carr_2017}: integrating Eq (\ref{eq:model_pde}) from $x = 0$ to $x = L$ and making use of Eq (\ref{eq:model_bcs}) yields:
\begin{align*}
\frac{\text{d}}{\text{d}t}\int_{0}^{L} T(x,t)\,\text{d}x = 0,
\end{align*}
and therefore $\int_{0}^{L} T(x,t)\,\text{d}x$ is constant for all $t$.} 

\revision{Integrating Eq (\ref{eq:u_ode_adiabatic}) twice with respect to $x$ and using the boundary and auxiliary conditions, Eq (\ref{eq:u_bcs_adiabatic})--(\ref{eq:u_constraint_adiabatic}), to identify the constants of integration yields the following solution of the} boundary value problem described by Eqs (\ref{eq:u_ode_adiabatic})--(\ref{eq:u_constraint_adiabatic}):
\begin{align*}
u(x) = \begin{cases} \displaystyle\frac{T_{\infty}L}{2\alpha}\left[\left(\frac{1}{\ell} - \frac{1}{L}\right)x^{2} - \frac{1}{3}\left(2L - 3\ell + \frac{\ell^{2}}{L}\right)\right],\\ \hspace*{5cm}\revision{0 \leq x \leq \ell},\\[0.2cm]
\displaystyle\frac{T_{\infty}L}{2\alpha}\left[- \frac{1}{L}x^{2} + 2x - \frac{1}{3}\left(2L+\frac{\ell^{2}}{L}\right)\right],\\ \hspace*{5cm}\revision{\ell \leq x \leq L}.\end{cases}
\end{align*}
Evaluating $u(x)$ at $x = L$ and recalling Eq (\ref{eq:u_integral_adiabatic}) yields:
\begin{gather}
\label{eq:integral_adiabatic}
\int_{0}^{\infty} \revision{\left[T_{\infty} - T_{r}(t)\right]}\,\text{d}t = \frac{T_{\infty}(L^{2}-\ell^{2})}{6\alpha},
\end{gather}
which can be rearranged to obtain the following formula for the thermal diffusivity 
\begin{gather}
\label{eq:alpha_carr}
\alpha = \frac{T_{\infty}(L^{2}-\ell^{2})}{6\int_{0}^{\infty} \revision{\left[T_{\infty} - T_{r}(t)\right]}\,\text{d}t}.
\end{gather}
Importantly, the above formula is exact. In contrast to Parker et al.'s formula, Eq (\ref{eq:alpha_parker}), we have made no approximations to arrive at Eq (\ref{eq:alpha_carr}). The equality in Eq (\ref{eq:alpha_parker}) remains true in the limit $\ell\rightarrow 0$ in which case 
\begin{gather}
\label{eq:alpha_carr_zero_depth}
\alpha = \frac{T_{\infty}L^{2}}{6\int_{0}^{\infty} \revision{\left[T_{\infty} - T_{r}(t)\right]}\,\text{d}t}.
\end{gather}
\revision{Eq (\ref{eq:alpha_carr_zero_depth}) can be derived from the form of $u(x)$ on $\ell\leq x\leq L$ which is valid for all $0\leq x \leq L$ in the limit as $\ell\rightarrow 0$. Evaluating $u(x)$ at $x = L$ and taking the limit as $\ell\rightarrow 0$ yields $u(L) = T_{\infty}L^{2}/(6\alpha)$ which when combined with Eq (\ref{eq:u_adiabatic}) gives Eq (\ref{eq:alpha_carr_zero_depth}).}

\section{Results}
\label{sec:results}
\begin{figure*}
\centering
\includegraphics{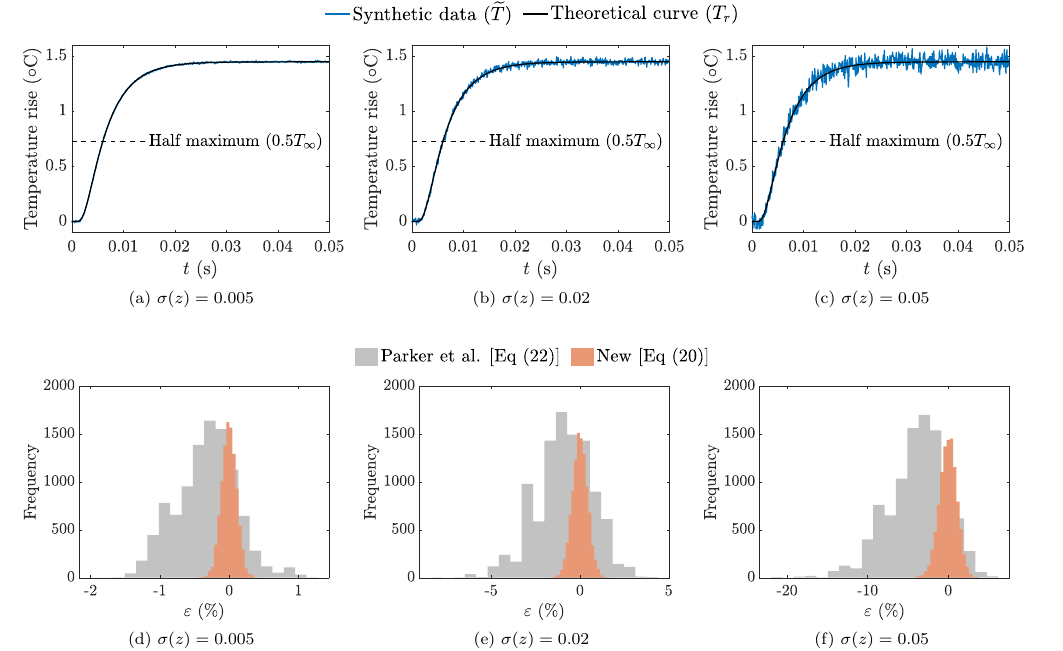}
\caption{(a)--(c) Plot of the theoretical rear-surface temperature rise curve, Eq (\ref{eq:Tr}) (black curves) and one example realisation of the synthetic rear-surface temperature rise data, Eq (\ref{eq:synethic_rear_surface_temperature}), for low ($\sigma(z) = 0.005$), moderate ($\sigma(z) = 0.02$) and high ($\sigma(z) = 0.05$) levels of noise (blue curves). \revision{Parameter values used to generate the synthetic data are given in Eqs (\ref{eq:parameters1})--(\ref{eq:parameters3})}. (d)--(f) Histograms of the signed relative error associated with the thermal diffusivity estimates obtained using Parker et al.'s formula, Eq (\ref{eq:alpha_parker_data}) (gray), and the new formula, Eq (\ref{eq:alpha_carr_data}) (orange), using 10,000 realisations of the synthetic rear-surface temperature rise data for each of the three different noise levels.}
\label{fig:results}
\end{figure*}

We now compare the accuracy of the new formula, Eq (\ref{eq:alpha_carr}), against Parker et al.'s formula, Eq (\ref{eq:alpha_parker}), for computing the thermal diffusivity. To perform the comparison, synthetic temperature rise data is generated at the rear-surface of the sample. This is achieved by adding Gaussian noise to the theoretical rear-surface temperature rise curve, Eq (\ref{eq:Tr}), at $N+1$ equally-spaced discrete times: 
\begin{gather}
\label{eq:synethic_rear_surface_temperature}
\widetilde{T}_{i} = T_{r}(t_{i}) + z_{i},\quad i = 0,\hdots, N,
\end{gather}
where $t_{i} = it_{N}/N$, $t_{N}$ is the final sampled time and $z_{i}$ is a random number sampled from a normal distribution with mean zero and standard deviation denoted by $\sigma(z)$ \revision{obtained using MATLAB's \texttt{randn} function}. In Eq (\ref{eq:synethic_rear_surface_temperature}), $T_{r}(t)$\revision{, Eq (\ref{eq:Tr}),} is evaluated at $t = t_{i}$ and the following set of parameter values \citep{czel_2013}: 
\revision{
\begin{gather}
\label{eq:parameters1}
L = 0.002\,\text{m},\quad Q = 7000\,\text{J}\,\text{m}^{-2},\\
\label{eq:parameters2}
k = 222\,\text{W}\,\text{m}^{-1}\text{K}^{-1},\quad \rho = 2700\,\text{kg}\,\text{m}^{-3},\\
\label{eq:parameters3}
c = 896\,\text{J}\,\text{kg}^{-1}\text{K}^{-1},\quad \ell = 0.0001\,\text{m},
\end{gather}}
\hspace*{-0.14cm}with the summation in Eq (\ref{eq:Tr}) truncated after $n = 200$ terms. The above parameter values lead to the following target value for the thermal diffusivity: 
\begin{gather}
\label{eq:target_diffusivity}
\alpha = \frac{k}{\rho c} = 9.1766\times 10^{-5}\,\text{m}^{2}\text{s}^{-1},
\end{gather}
rounded to five significant figures. Figures \ref{fig:results}(a)--(c) depict example synthetic rear-surface temperature rise data for three noise levels: low ($\sigma(z) = 0.005\,^{\circ}\text{C}$), moderate ($\sigma(z) = 0.02\,^{\circ}\text{C}$) and high ($\sigma(z) = 0.05\,^{\circ}\text{C}$) \citep{czel_2013}. 

To estimate the thermal diffusivity according to Parker et al.'s formula, Eq (\ref{eq:alpha_parker}), the value of the half rise time $t_{0.5}$ must be estimated from the  temperature rise data, Eq (\ref{eq:synethic_rear_surface_temperature}). Similarly, for the newly derived formula, Eq (\ref{eq:alpha_carr}), one must approximate the integral appearing in the denominator. The latter approximation is carried out using the trapezoidal rule, giving:
\begin{gather}
\label{eq:alpha_carr_data}
\alpha \approx \frac{(L^{2}-\ell^{2})}{6}\left[\sum_{i=1}^{N}\left(1 - \frac{\widetilde{T}_{i-1}+\widetilde{T}_{i}}{2T_{\infty}}\right)\Delta t_{i}\right]^{-1},
\end{gather}
where $\Delta t_{i} = t_{i}-t_{i-1}$. To ensure the integral in Eq (\ref{eq:alpha_carr}) is well approximated, $t_{N}$ must be sufficiently large so that $\widetilde{T}_{N}$ is close to $T_{\infty}$. \revision{Once the thermal diffusivity has been estimated from Eq (\ref{eq:alpha_carr_data}), the appropriateness of the value of $t_{N}$ can be verified by calculating the time required for $T_{r}(t)$, Eq (\ref{eq:Tr}), to transition to within a prescribed tolerance, $\delta$, of $T_{\infty}$. This time is known as the finite transition time \citep{carr_2017} or steady-state time \citep{dalessandro_2018} and can be estimated by truncating the summation in Eq (\ref{eq:Tr}) after $n= 1$ term.} 

In this work, the half rise time in Eq (\ref{eq:alpha_parker}) is calculated using linear interpolation:
\begin{gather}
t_{0.5} = t_{j-1} + \frac{0.5T_{\infty} - \widetilde{T}_{j-1}}{\widetilde{T}_{j}-\widetilde{T}_{j-1}}\Delta t_{j},
\end{gather}
where $j$ is the smallest index (corresponding to the smallest discrete time $t_{j}$) satisfying $\widetilde{T}_{j} > 0.5T_{\infty}$, giving:
\begin{gather}
\label{eq:alpha_parker_data}
\alpha \approx \frac{\omega_{0.5}L^{2}}{\pi^{2}}\left[t_{j-1} + \frac{0.5T_{\infty} - \widetilde{T}_{j-1}}{\widetilde{T}_{j}-\widetilde{T}_{j-1}}\Delta t_{j}\right]^{-1}.
\end{gather}
Histograms and corresponding summary statistics comparing the two formulae for estimating the thermal diffusivity, Eqs (\ref{eq:alpha_carr_data}) and (\ref{eq:alpha_parker_data}), are given in Figure \ref{fig:results} and Table \ref{tab:results1}. The results are based on 10,000 realisations of synthetic rear-surface temperature rise data generated according to Eq (\ref{eq:synethic_rear_surface_temperature}) with $t_{N} = 0.05\,\text{s}$ and $N = 500$ \citep{czel_2013}. For each realisation, estimates of the thermal diffusivity, denoted $\widetilde{\alpha}$, are calculated according to Eqs (\ref{eq:alpha_carr_data}) and (\ref{eq:alpha_parker_data}) and compared to the target value given in Eq (\ref{eq:target_diffusivity}) by computing the signed relative error $\varepsilon = (\alpha - \widetilde{\alpha})/\alpha \times 100\%$.  All values for the signed relative error and thermal diffusivity in Table \ref{tab:results1} are reported to one and five significant figures, respectively.

\begin{table*}
\centering
\begin{tabular}{|l|r|r|r|r|r|r|r|r|}
\hline
Formula & $\sigma(z)$ & $\mu(\varepsilon)$ & $\sigma(\varepsilon)$ & $\min(\varepsilon)$ & $\max(\varepsilon)$ & $\mu(\widetilde{\alpha})$ & $\min(\widetilde{\alpha})$ & $\max(\widetilde{\alpha})$\\
\hline & & & & & & & &\\[-0.35cm]
Parker et al. & 0.005 & \num{-0.3} & \num{0.4} & \num{-2} & \num{1} & \num{9.2069e-05} & \num{9.0609e-05} & \num{9.3554e-05}\\
{[}Eq (\ref{eq:alpha_parker_data}){]} & 0.02 & \num{-0.9} & \num{2} & \num{-8} & \num{4} & \num{9.2573e-05} & \num{8.7818e-05} & \num{9.9239e-05}\\
& 0.05 & \num{-4} & \num{4} & \num{-20} & \num{6} & \num{9.5282e-05} & \num{8.6096e-05} & \num{1.1124e-04}\\
\hline & & & & & & & &\\[-0.35cm]
New & 0.005 & \num{-0.001} & \num{0.1} & \num{-0.5} & \num{0.4} & \num{9.1767e-05} & \num{9.1406e-05} & \num{9.2183e-05}\\
{[}Eq (\ref{eq:alpha_carr_data}){]} & 0.02 & \num{0.002} & \num{0.4} & \num{-2} & \num{2} & \num{9.1764e-05} & \num{9.0305e-05} & \num{9.3162e-05}\\
& 0.05 & \num{-0.02} & \num{1} & \num{-4} & \num{4} & \num{9.1781e-05} & \num{8.8409e-05} & \num{9.5515e-05}\\
\hline
\end{tabular}
\caption{Summary statistics for the histograms in Figure \ref{fig:results}(d)--(f). Statistics in this table include the standard deviation of the noise, $\sigma(z)$, the mean and standard deviation of the signed relative error, $\mu(\varepsilon)$ and $\sigma(\varepsilon)$, the minimum and maximum values of the signed relative error, $\min(\varepsilon)$ and $\max(\varepsilon)$, the mean thermal diffusivity estimate, $\mu(\widetilde{\alpha})$, and the minimum and maximum values of the thermal diffusivity estimate, $\min(\widetilde{\alpha})$ and $\max(\widetilde{\alpha})$.}
\label{tab:results1}
\end{table*}

\begin{table*}
\centering
\revision{
\begin{tabular}{|l|r|r|r|r|r|r|r|r|}
\hline
Formula & $\sigma(z)$ & $\mu(\varepsilon)$ & $\sigma(\varepsilon)$ & $\min(\varepsilon)$ & $\max(\varepsilon)$ & $\mu(\widetilde{\alpha})$ & $\min(\widetilde{\alpha})$ & $\max(\widetilde{\alpha})$\\
\hline & & & & & & & &\\[-0.35cm]
Parker et al. & 0.005 & \num{0.006} & \num{0.5} & \num{-1} & \num{2} & \num{9.1760e-05} & \num{9.0363e-05} & \num{9.3044e-05}\\
{[}Eq (\ref{eq:alpha_parker_data}){]} & 0.02 & \num{-0.5} & \num{2} & \num{-7} & \num{4} & \num{9.2268e-05} & \num{8.7839e-05} & \num{9.8068e-05}\\
& 0.05 & \num{-4} & \num{4} & \num{-20} & \num{6} & \num{9.5018e-05} & \num{8.5876e-05} & \num{1.1331e-04}\\
\hline & & & & & & & &\\[-0.35cm]
New & 0.005 & \num{-0.0003} & \num{0.1} & \num{-0.4} & \num{0.4} & \num{9.1766e-05} & \num{9.1407e-05} & \num{9.2162e-05}\\
{[}Eq (\ref{eq:alpha_carr_data}){]} & 0.02 & \num{-0.007} & \num{0.4} & \num{-2} & \num{2} & \num{9.1772e-05} & \num{9.0318e-05} & \num{9.3310e-05}\\
& 0.05 & \num{-0.02} & \num{1} & \num{-4} & \num{4} & \num{9.1781e-05} & \num{8.8242e-05} & \num{9.5151e-05}\\
\hline
\end{tabular}
\caption{Summary statistics for the thermal diffusivity estimates arising from 10,000 realisations of the synthetic rear-surface temperature rise data for low ($\sigma(z) = 0.005$), moderate ($\sigma(z) = 0.02$) and high ($\sigma(z) = 0.05$) levels of noise. \revision{Parameter values used to generate the synthetic data are given in Eqs (\ref{eq:parameters1})--(\ref{eq:parameters3}) except with $\ell = 0$}. Notation used in this table is defined in Table \ref{tab:results1}.}
\label{tab:results2}}
\end{table*}

\begin{table*}
\centering
\revision{
\begin{tabular}{|l|r|r|r|r|r|r|r|r|r|}
\hline
$\sigma(\ell)$ & $\sigma(z)$ & $\mu(\varepsilon)$ & $\sigma(\varepsilon)$ & $\min(\varepsilon)$ & $\max(\varepsilon)$ & $\mu(\widetilde{\alpha})$ & $\min(\widetilde{\alpha})$ & $\max(\widetilde{\alpha})$\\
\hline & & & & & & & &\\[-0.35cm]
$\num{5e-06}$ & 0.005 & \num{-0.0006} & \num{0.1} & \num{-0.5} & \num{0.4} & \num{9.1766e-05} & \num{9.1412e-05} & \num{9.2211e-05}\\
& 0.02 & \num{0.003} & \num{0.4} & \num{-2} & \num{2} & \num{9.1763e-05} & \num{9.0285e-05} & \num{9.3187e-05}\\
& 0.05 & \num{-0.02} & \num{1} & \num{-4} & \num{4} & \num{9.1781e-05} & \num{8.8391e-05} & \num{9.5510e-05}\\
\hline & & & & & & & &\\[-0.35cm]
$\num{5e-05}$ & 0.005 & \num{0.06} & \num{0.3} & \num{-0.6} & \num{2} & \num{9.1710e-05} & \num{9.0184e-05} & \num{9.2311e-05}\\
& 0.02 & \num{0.06} & \num{0.5} & \num{-2} & \num{2} & \num{9.1712e-05} & \num{8.9675e-05} & \num{9.3325e-05}\\
& 0.05 & \num{0.05} & \num{1} & \num{-4} & \num{4} & \num{9.1722e-05} & \num{8.7957e-05} & \num{9.5555e-05}\\
\hline
\end{tabular}
\caption{Summary statistics for thermal diffusivity estimates calculated using Eq (\ref{eq:alpha_carr_data}) from 10,000 realisations of the synthetic rear-surface temperature rise data (same realisations as Table \ref{tab:results1}). To account for possible experimental error in $\ell$, Gaussian noise of mean zero and standard deviation $\sigma(\ell)$ is added to the value of $\ell$ used in Eq (\ref{eq:alpha_carr_data}). Results are shown for two standard deviations, $\sigma(\ell) = 10^{-6}\,\text{m}$ and $\sigma(\ell) = 10^{-4}\,\text{m}$, each combined with the low ($\sigma(z) = 0.005$), moderate ($\sigma(z) = 0.02$) and high ($\sigma(z) = 0.05$) levels of noise imposed on the synthetic rear-surface temperature rise data. Notation used in this table is defined in Table \ref{tab:results1}.}
\label{tab:results3}}
\end{table*}

\begin{table*}
\centering
\revision{
\begin{tabular}{|r|r|r|r|r|r|r|r|r|}
\hline
$\sigma(z)$ & $\mu(\varepsilon)$ & $\sigma(\varepsilon)$ & $\min(\varepsilon)$ & $\max(\varepsilon)$ & $\mu(\widetilde{\alpha})$ & $\min(\widetilde{\alpha})$ & $\max(\widetilde{\alpha})$\\
\hline & & & & & & &\\[-0.35cm]
0.005 & \num{-0.3} & \num{0.1} & \num{-0.7} & \num{0.1} & \num{9.1997e-05} & \num{9.1635e-05} & \num{9.2414e-05}\\
0.02 & \num{-0.2} & \num{0.4} & \num{-2} & \num{1} & \num{9.1994e-05} & \num{9.0531e-05} & \num{9.3395e-05}\\
0.05 & \num{-0.3} & \num{1} & \num{-4} & \num{3} & \num{9.2011e-05} & \num{8.8630e-05} & \num{9.5754e-05}\\
\hline
\end{tabular}
\caption{Summary statistics for thermal diffusivity estimates calculated using Eq (\ref{eq:alpha_carr_data}) with $\ell = 0$ from 10,000 realisations of the synthetic rear-surface temperature rise data (same realisations as Table \ref{tab:results1}). Results are shown for low ($\sigma(z) = 0.005$), moderate ($\sigma(z) = 0.02$) and high ($\sigma(z) = 0.05$) levels of noise. Notation used in this table is defined in Table \ref{tab:results2}. Notation used in this table is defined in Table \ref{tab:results1}.}
\label{tab:results4}}
\end{table*}

The key observation from these results is that for all three noise levels, the new formula, Eq (\ref{eq:alpha_carr_data}),  produces more accurate and less variable estimates of the thermal diffusivity. This is clearly evident by the location and spread of the histograms in Figures \ref{fig:results}(d)--(f) as well as in Table \ref{tab:results1} where the implementation of Parker et al.'s formula results in significantly larger values for the mean and standard deviation of the signed relative error across all three noise levels. For the chosen test case, Parker et al.'s formula, Eq (\ref{eq:alpha_parker_data}), is accurate to within $-2\%$ and $+1\%$ for the low level of noise, $-8\%$ and $+4\%$ for the moderate level of noise and $-20\%$ and $+6\%$ for the high level of noise. Comparatively, the new formula, Eq (\ref{eq:alpha_carr_data}), is accurate to within $-0.5\%$ and $+0.4\%$, $-2\%$ and $+2\%$, and $-4\%$ and $+4\%$ for the low, moderate and high levels of noise, respectively. Comparing the mean estimates of the thermal diffusivity in Table \ref{tab:results1} it is clear that across all three noise levels the new formula, Eq (\ref{eq:alpha_carr_data}), provides superior agreement with the target value given in Eq (\ref{eq:target_diffusivity}). 

It is worth noting that the estimates of the thermal diffusivity obtained using Parker et al.'s formula can be improved by first smoothing the data. To explore this further, an interesting experiment is to apply Eq (\ref{eq:alpha_parker_data}) to the best possible smoothed data, that is, Eq (\ref{eq:synethic_rear_surface_temperature}) with zero noise ($z_{i} = 0$ for all $i = 0,\hdots,N$). This exercise yields a thermal diffusivity estimate of $9.2039\times 10^{-5}$ and a signed relative error of $-0.3\%$. \revisiontwo{This outcome and the inferior performance of Parker et al.'s formula in Table \ref{tab:results1} suggests that there is an inherent bias in the half-rise time approach that is most likely attributed to the fact that the analysis used to derive the estimate of the thermal diffusivity is based on the approximate rear-surface temperature rise curve obtained in the limit $\ell\rightarrow 0$, Eq (\ref{eq:Tr}), rather than the true curve, Eq (\ref{eq:rear_surface_temperature_rise_approx}).} We now therefore repeat the same experiment carried out in Figure \ref{fig:results} and Table \ref{tab:results1} except with $\ell = 0$. In this case, we have two exact expressions for the thermal diffusivity: the new formula, Eq (\ref{eq:alpha_carr}), reduces to Eq (\ref{eq:alpha_carr_zero_depth}) while the approximately equal sign becomes an equality in Parker et al.'s formula, Eq (\ref{eq:alpha_parker}). Nevertheless, both methods incur error due to the approximations of the rear-surface integral and the half rise time required in Eqs (\ref{eq:alpha_carr_data}) and (\ref{eq:alpha_parker_data}), respectively. As expected, the accuracy of Parker et al.'s formula is better for $\ell = 0$ (Table \ref{tab:results2}) than for $\ell = 0.0001\,\text{m}$ (Table \ref{tab:results1}). However, across all three noise levels the new formula, Eq (\ref{eq:alpha_carr_data}), still provides superior agreement with the target value given in Eq (\ref{eq:target_diffusivity}). 

A disadvantage of the new formula, Eq (\ref{eq:alpha_carr_data}), is that it requires $\ell$, the depth at the front surface of the sample in which the heat pulse is absorbed, which must be measured and is therefore susceptible to measurement error. To account for this in \revisiontwo{the} analysis we add Gaussian noise of mean zero and standard deviation $\sigma(\ell)$ to the value of $\ell = 0.0001\,\text{m}$ used in Eq (\ref{eq:alpha_carr_data}). The perturbed value of $\ell$ does not influence the synthetic rear-surface temperature rise data, Eq (\ref{eq:synethic_rear_surface_temperature}), nor Parker et al.'s formula, Eq (\ref{eq:alpha_parker_data}), where $\ell$ is absent. Using the same 10,000 realisations of the synthetic rear-surface temperature rise data from Table \ref{tab:results1}, we give results in Table \ref{tab:results3} for two standard deviations, $\sigma(\ell) = 5\times10^{-6}\,\text{m}$ and $\sigma(\ell) = 5\times 10^{-5}\,\text{m}$, each combined with the three standard deviations imposed on the synthetic rear-surface temperature rise data previously considered. To avoid possible non-physical values of $\ell$ we take the value of $\ell$ as the maximum of the perturbed value and zero. Results in Table \ref{tab:results3} demonstrate that even with noise added to the depth $\ell$, in all cases, the mean and standard deviation of the signed relative error for the new formula, Eq (\ref{eq:alpha_carr_data}), remains smaller than the corresponding values for Parker et al.'s formula, Eq (\ref{eq:alpha_parker_data})  from Table \ref{tab:results1}.

An interesting exercise is to neglect the value of $\ell$ in Eq (\ref{eq:alpha_carr_data}) and use the resulting formula, equivalent to the one obtained by applying the trapezoidal rule to Eq (\ref{eq:alpha_carr_zero_depth}), to calculate the thermal diffusivity. The outcome of this exercise is shown in Table \ref{tab:results4} using the same 10,000 realisations of the synthetic rear-surface temperature rise data from Table \ref{tab:results1}, which are generated using $\ell = 0.0001\,\text{m}$. Compared to the results presented in Table \ref{tab:results1}, the new formula remains more accurate and less variable than Parker et al.'s formula, Eq (\ref{eq:alpha_parker_data}).

\section{Conclusion}
\label{sec:conclusions}
In conclusion, this paper has presented a novel method for calculating thermal diffusivity from laser flash experiments. The proposed method expresses the thermal diffusivity exactly in terms of the area enclosed by the rear-surface temperature rise curve and the steady-state temperature over time. For a typical test case, \revision{the new} method produced estimates of the thermal diffusivity that are more accurate and less variable than the standard half-rise time approach of \citet{parker_1961}.

\revision{It is important to remember that the thermal diffusivity formulae, Eqs (\ref{eq:alpha_carr}) and (\ref{eq:alpha_carr_data}), apply only in the case of ideal conditions (as described in the introduction).} \revisiontwo{These conditions include the assumption of a uniform initial distribution of heat in the layer in which the heat pulse is absorbed, Eq (\ref{eq:model_ic}), and the assumption of a thermally-insulated sample, Eq (\ref{eq:model_bcs}). While future work will focus on overcoming the limitation of the first assumption, extension to the case of heat losses \citep{parker_1962,cowan_1963}, is straightforward}. Under the assumption that the ambient temperature and initial temperature of the sample are equal, the boundary conditions at the front and rear surface of the sample, Eq (\ref{eq:model_bcs}), become:
\begin{gather}
\begin{split}
\label{eq:model_bcs_loss}
\frac{\partial T}{\partial x}(0,t) - h_{0}T(0,t) = 0,\\ \frac{\partial T}{\partial x}(L,t) + h_{L}T(L,t) = 0,
\end{split}
\end{gather}
where $h_{0}$ and $h_{L}$ are linear heat transfer coefficients scaled by the thermal conductivity \citep{parker_1962}. With Eq (\ref{eq:model_bcs_loss}), the steady state solution of Eqs (\ref{eq:model_pde}), (\ref{eq:model_ic}) and (\ref{eq:model_bcs_loss}) is zero for all values of $x$. In this case, the area enclosed by the rear-surface temperature rise curve, $T_{r}(t)$, and the steady state temperature, $T_{\infty}=0$, between $t = 0$ and $t\rightarrow\infty$ is expressed as $\int_{0}^{\infty} T_{r}(t)\,\text{d}t$. Following a similar procedure to the one outlined earlier for the adiabatic case yields the closed-form expression
\begin{gather}
\label{eq:integral_loss}
\int_{0}^{\infty} T_{r}(t)\,\text{d}t = \frac{Q(\ell h_{0}+2)}{2\alpha \rho c(Lh_{0}h_{L}+h_{0}+h_{L})}.
\end{gather}
Rearranging Eq (\ref{eq:integral_loss}) produces the following formula for the thermal diffusivity 
\begin{gather}
\label{eq:alpha_carr_loss}
\alpha = \frac{Q(\ell h_{0}+2)}{2 \rho c(Lh_{0}h_{L}+h_{0}+h_{L})}\left[\int_{0}^{\infty} T_{r}(t)\,\text{d}t\right]^{-1},
\end{gather}
with application of the trapezoidal rule yielding the equivalent of Eq (\ref{eq:alpha_carr_data}) for the case of heat losses:
\begin{gather*}
\alpha \approx \frac{Q(\ell h_{0}+2)}{2 \rho c(Lh_{0}h_{L}+h_{0}+h_{L})}\left[\sum_{i=1}^{N} \frac{\widetilde{T}_{i-1}+\widetilde{T}_{i}}{2}\Delta t_{i}\right]^{-1}.
\end{gather*}
\revisiontwo{Eq (\ref{eq:alpha_carr_loss}) provides a formula for the thermal diffusivity for a sample of known volumetric heat capacity, $\rho c$. If the volumetric heat capacity is unknown then dividing both sides of Eq (\ref{eq:alpha_carr_loss}) by $\rho c$ yields a formula for the thermal conductivity, another much sought after property:
\begin{gather}
\label{eq:k_carr_loss}
k = \frac{Q(\ell h_{0}+2)}{2(Lh_{0}h_{L}+h_{0}+h_{L})}\left[\int_{0}^{\infty} T_{r}(t)\,\text{d}t\right]^{-1}.
\end{gather}
Moreover, Eq (\ref{eq:alpha_carr_loss}) may be useful when curve-fitting methods are applied to calculate thermo-physical properties \citep{gembarovic_1994} as it can be used to eliminate either thermal diffusivity or volumetric heat capacity from the set of parameters that are estimated in the curve-fitting procedure.}

\section*{Acknowledgements}
This research was funded by the Australian Research Council (DE150101137). \revisiontwo{Insightful comments from two anonymous referees greatly improved the quality of this paper.}


\bibliographystyle{model2-names}
\bibliography{references}

\begin{thebibliography}{22}
\expandafter\ifx\csname natexlab\endcsname\relax\def\natexlab#1{#1}\fi
\providecommand{\url}[1]{\texttt{#1}}
\providecommand{\href}[2]{#2}
\providecommand{\path}[1]{#1}
\providecommand{\DOIprefix}{doi:}
\providecommand{\ArXivprefix}{arXiv:}
\providecommand{\URLprefix}{URL: }
\providecommand{\Pubmedprefix}{pmid:}
\providecommand{\doi}[1]{\href{http://dx.doi.org/#1}{\path{#1}}}
\providecommand{\Pubmed}[1]{\href{pmid:#1}{\path{#1}}}
\providecommand{\bibinfo}[2]{#2}
\ifx\xfnm\relax \def\xfnm[#1]{\unskip,\space#1}\fi
\bibitem[{{ASTM E1461-13}(2013)}]{ASTM_E1461}
\bibinfo{author}{{ASTM E1461-13}}, \bibinfo{year}{2013}.
\newblock \bibinfo{title}{Standard test method for thermal diffusivity by the
  flash method}.
\newblock \bibinfo{journal}{West Conshohocken, PA} .
\bibitem[{Azumi and Takahashi(1981)}]{azumi_1981}
\bibinfo{author}{Azumi, T.}, \bibinfo{author}{Takahashi, Y.},
  \bibinfo{year}{1981}.
\newblock \bibinfo{title}{Novel finite pulse-wide correction in flash thermal
  diffusivity measurement}.
\newblock \bibinfo{journal}{Rev. Sci. Instrum.} \bibinfo{volume}{52},
  \bibinfo{pages}{1411--1413}.
\bibitem[{Blumm and Opfermann(2002)}]{blumm_2002}
\bibinfo{author}{Blumm, J.}, \bibinfo{author}{Opfermann, J.},
  \bibinfo{year}{2002}.
\newblock \bibinfo{title}{Improvement of the mathematical modelling of flash
  measurements}.
\newblock \bibinfo{journal}{High Temp.-High Press.} \bibinfo{volume}{34},
  \bibinfo{pages}{515--521}.
\bibitem[{Cape and Lehman(1963)}]{cape_1963}
\bibinfo{author}{Cape, J.A.}, \bibinfo{author}{Lehman, G.W.},
  \bibinfo{year}{1963}.
\newblock \bibinfo{title}{Temperature and finite pulse-time effects in the
  flash method for measuring thermal diffusivity}.
\newblock \bibinfo{journal}{J. Appl. Phys.} \bibinfo{volume}{34},
  \bibinfo{pages}{1909--1913}.
\bibitem[{Carr(2017)}]{carr_2017}
\bibinfo{author}{Carr, E.J.}, \bibinfo{year}{2017}.
\newblock \bibinfo{title}{Calculating how long it takes for a diffusion process
  to effectively reach steady state without computing the transient solution}.
\newblock \bibinfo{journal}{Phys. Rev. E} \bibinfo{volume}{96},
  \bibinfo{pages}{012116}.
\bibitem[{Carr(2018)}]{carr_2018}
\bibinfo{author}{Carr, E.J.}, \bibinfo{year}{2018}.
\newblock \bibinfo{title}{Characteristic timescales for diffusion processes
  through layers and across interfaces}.
\newblock \bibinfo{journal}{Phys. Rev. E} \bibinfo{volume}{97},
  \bibinfo{pages}{042115}.
\bibitem[{Chen et~al.(2010)Chen, Limarga and Clarke}]{chen_2010}
\bibinfo{author}{Chen, L.}, \bibinfo{author}{Limarga, A.M.},
  \bibinfo{author}{Clarke, D.R.}, \bibinfo{year}{2010}.
\newblock \bibinfo{title}{A new data reduction method for pulse diffusivity
  measurements on coated samples}.
\newblock \bibinfo{journal}{Comp. Mater. Sci} \bibinfo{volume}{50},
  \bibinfo{pages}{77--82}.
\bibitem[{Cowan(1963)}]{cowan_1963}
\bibinfo{author}{Cowan, R.D.}, \bibinfo{year}{1963}.
\newblock \bibinfo{title}{Pulse method for measuring thermal diffusivity at
  high temperatures}.
\newblock \bibinfo{journal}{J. Appl. Phys.} \bibinfo{volume}{34},
  \bibinfo{pages}{926--927}.
\bibitem[{Cz{\'e}l et~al.(2013)Cz{\'e}l, Woodbury, Woolley and
  Gr{\'o}f}]{czel_2013}
\bibinfo{author}{Cz{\'e}l, B.}, \bibinfo{author}{Woodbury, K.A.},
  \bibinfo{author}{Woolley, J.}, \bibinfo{author}{Gr{\'o}f, G.},
  \bibinfo{year}{2013}.
\newblock \bibinfo{title}{Analysis of parameter estimation possibilities of the
  thermal contact resistance using the laser flash method with two-layer
  specimens}.
\newblock \bibinfo{journal}{Int. J. Thermophys.} \bibinfo{volume}{34},
  \bibinfo{pages}{1993--2008}.
\bibitem[{D'Alessandro and de~Monte(2018)}]{dalessandro_2018}
\bibinfo{author}{D'Alessandro, G.}, \bibinfo{author}{de~Monte, F.},
  \bibinfo{year}{2018}.
\newblock \bibinfo{title}{Intrinsic verification of an exact analytical
  solution in transient heat conduction}.
\newblock \bibinfo{journal}{Computational Thermal Sciences}
  \bibinfo{volume}{10}, \bibinfo{pages}{251--272}.
\bibitem[{Ellery et~al.(2012)Ellery, Simpson, McCue and Baker}]{ellery_2012b}
\bibinfo{author}{Ellery, A.J.}, \bibinfo{author}{Simpson, M.J.},
  \bibinfo{author}{McCue, S.W.}, \bibinfo{author}{Baker, R.E.},
  \bibinfo{year}{2012}.
\newblock \bibinfo{title}{Critical time scales for advection-diffusion-reaction
  processes}.
\newblock \bibinfo{journal}{Phys. Rev. E} \bibinfo{volume}{85},
  \bibinfo{pages}{041135}.
\bibitem[{Gembarovi\u{c} and Taylor(1994)}]{gembarovic_1994}
\bibinfo{author}{Gembarovi\u{c}, J.}, \bibinfo{author}{Taylor, R.E.},
  \bibinfo{year}{1994}.
\newblock \bibinfo{title}{A new data reduction procedure in the flash method of
  measuring thermal diffusivity}.
\newblock \bibinfo{journal}{Rev. Sci. Instrum.} \bibinfo{volume}{65},
  \bibinfo{pages}{3535--3539}.
\bibitem[{Heckman(1973)}]{heckman_1973}
\bibinfo{author}{Heckman, R.C.}, \bibinfo{year}{1973}.
\newblock \bibinfo{title}{Finite pulse-time and heat-loss effets in pulse
  thermal diffusivity measurements}.
\newblock \bibinfo{journal}{J. Appl. Phys.} \bibinfo{volume}{44},
  \bibinfo{pages}{1455--1460}.
\bibitem[{James(1980)}]{james_1980}
\bibinfo{author}{James, H.M.}, \bibinfo{year}{1980}.
\newblock \bibinfo{title}{Some extensions of the flash method of measuring
  thermal diffusivity}.
\newblock \bibinfo{journal}{J. Appl. Phys.} \bibinfo{volume}{51},
  \bibinfo{pages}{4666--4672}.
\bibitem[{Josell et~al.(1995)Josell, Warren and Cezairliyan}]{josell_1995}
\bibinfo{author}{Josell, D.}, \bibinfo{author}{Warren, J.},
  \bibinfo{author}{Cezairliyan, A.}, \bibinfo{year}{1995}.
\newblock \bibinfo{title}{Comment on ``analysis for determining thermal
  diffusivity from thermal pulse experiments''}.
\newblock \bibinfo{journal}{J. Appl. Phys.} \bibinfo{volume}{78},
  \bibinfo{pages}{6867--6869}.
\bibitem[{Parker and Jenkins(1962)}]{parker_1962}
\bibinfo{author}{Parker, W.J.}, \bibinfo{author}{Jenkins, R.J.},
  \bibinfo{year}{1962}.
\newblock \bibinfo{title}{Thermal conductivity measurements on bismuth
  telluride in the presence of a 2 {MeV} electron beam}.
\newblock \bibinfo{journal}{Adv. Energy Conversion} \bibinfo{volume}{2},
  \bibinfo{pages}{87--103}.
\bibitem[{Parker et~al.(1961)Parker, Jenkins, Butler and Abbott}]{parker_1961}
\bibinfo{author}{Parker, W.J.}, \bibinfo{author}{Jenkins, R.J.},
  \bibinfo{author}{Butler, C.P.}, \bibinfo{author}{Abbott, G.L.},
  \bibinfo{year}{1961}.
\newblock \bibinfo{title}{Flash method of determining thermal diffusivity, heat
  capacity, and thermal conductivity}.
\newblock \bibinfo{journal}{J. Appl. Phys.} \bibinfo{volume}{32},
  \bibinfo{pages}{1679--1684}.
\bibitem[{Simpson et~al.(2013)Simpson, Jazaei and Clement}]{simpson_2013}
\bibinfo{author}{Simpson, M.J.}, \bibinfo{author}{Jazaei, F.},
  \bibinfo{author}{Clement, T.P.}, \bibinfo{year}{2013}.
\newblock \bibinfo{title}{How long does it take for aquifer recharge or aquifer
  discharge processes to reach steady state?}
\newblock \bibinfo{journal}{J. Hydrology} \bibinfo{volume}{501},
  \bibinfo{pages}{241--248}.
\bibitem[{Tao et~al.(2015)Tao, Yang, Zhong, Xu and Luo}]{tao_2015}
\bibinfo{author}{Tao, Y.}, \bibinfo{author}{Yang, L.}, \bibinfo{author}{Zhong,
  Q.}, \bibinfo{author}{Xu, Z.}, \bibinfo{author}{Luo, C.},
  \bibinfo{year}{2015}.
\newblock \bibinfo{title}{A modified evaluation method to reduce finite pulse
  time effects in flash diffusivity measurement}.
\newblock \bibinfo{journal}{Rev. Sci. Instrum.} \bibinfo{volume}{86},
  \bibinfo{pages}{124902}.
\bibitem[{Taylor and Cape(1964)}]{taylor_1964}
\bibinfo{author}{Taylor, R.E.}, \bibinfo{author}{Cape, J.A.},
  \bibinfo{year}{1964}.
\newblock \bibinfo{title}{Finite pulse-time effects in the flash diffusivity
  technique}.
\newblock \bibinfo{journal}{Appl. Phys. Lett.} \bibinfo{volume}{5},
  \bibinfo{pages}{212--213}.
\bibitem[{Voz{\'a}r and Hohenauer(2003)}]{vozar_2003}
\bibinfo{author}{Voz{\'a}r, L.}, \bibinfo{author}{Hohenauer, W.},
  \bibinfo{year}{2003}.
\newblock \bibinfo{title}{Flash method of measuring the thermal diffusivity. a
  review}.
\newblock \bibinfo{journal}{High Temp.-High Press.} \bibinfo{volume}{35-36},
  \bibinfo{pages}{253--264}.
\bibitem[{Zhao et~al.(2016)Zhao, Wu and Bai}]{zhao_2016}
\bibinfo{author}{Zhao, Y.H.}, \bibinfo{author}{Wu, Z.K.}, \bibinfo{author}{Bai,
  S.L.}, \bibinfo{year}{2016}.
\newblock \bibinfo{title}{Thermal resistance measurement of 3d graphene
  foam/polymer composite by laser flash analysis}.
\newblock \bibinfo{journal}{Int. J. Heat Mass Tran.} \bibinfo{volume}{101},
  \bibinfo{pages}{470--475}.

\end{thebibliography}

\end{document}